\documentclass{ws-procs9x6}

\newcommand{\be}{\begin{eqnarray}}
\newcommand{\ee}{\end{eqnarray}}
\newcommand{\non}{\nonumber}

\begin{document}

\title{Integrability + Supersymmetry + Boundary: \\
Life on the edge is not so dull after all!}

\author{Rafael I. Nepomechie\footnote{\uppercase{W}ork supported in
part by the \uppercase{N}ational \uppercase{S}cience
\uppercase{F}oundation under \uppercase{G}rants
\uppercase{PHY}-0098088 and \uppercase{PHY}-0244261.}}

\address{Physics Department, P.O. Box 248046, University of Miami\\
Coral Gables, FL 33124, USA\\
E-mail: nepomechie@physics.miami.edu}

\maketitle

\abstracts{
After a brief review of integrability, first in the absence and then
in the presence of a boundary, I outline the construction of actions
for the $N=1$ and $N=2$ boundary sine-Gordon models.  The key point is
to introduce Fermionic boundary degrees of freedom in the boundary
actions.}

\section{Introduction}\label{sec:intro}

Quantum field theories (QFTs) with enhanced spacetime symmetries, such
as integrability or supersymmetry, are attractive to theorists both as
candidate models of real physical systems, and as toy models which can
be probed more deeply than would otherwise be possible by exploiting
their symmetries.  Introducing a spatial boundary in such theories,
whose effects can be physically important, poses a particular
challenge to theorists, since boundary conditions generically break
bulk spacetime symmetries.

Hence the fundamental question: to what extent can bulk spacetime
symmetries be maintained in the presence of a spatial boundary?  One
expects that the ``more'' bulk symmetries there are, the harder it is
to maintain such symmetries when a boundary is introduced.  In
particular, it was believed for some years that it is essentially
impossible to maintain both integrability and supersymmetry in the
presence of a boundary.  

My main message here is that this belief is wrong: there {\bf do}
exist nontrivial integrable supersymmetric boundary QFTs.  Although I
address this question in the specific context of the sine-Gordon
model, I expect that corresponding results can also be found for other
models.  This result may have applications in various areas, including
condensed matter physics (in connection with impurity problems) and
superstring theory.  However, I have been motivated not so much by any
particular application, but rather, by the two general convictions
that systems with enhanced spacetime symmetries can be very
interesting, and that boundary effects can be very important.

The remainder of this article is organized as follows.  Sec.  2
briefly reviews some general features of integrability in the absence
of boundaries, and considers as an example the sine-Gordon model
\cite{ZZ}.  Sec.  3 briefly reviews integrability in the presence of a
boundary, focusing on the boundary sine-Gordon model \cite{GZ}.  Secs.
4 and 5 outline the construction of actions for the $N=1$ and $N=2$
boundary sine-Gordon models, respectively \cite{Ne1,Ne2}.  The key
point is to introduce Fermionic boundary degrees of freedom in the
boundary actions.  Sec.  6 lists some interesting open problems, and
points out related recent work on superconformal boundary Liouville
models.

\section{Integrability}\label{sec:integrability}

In this Section, I very briefly review some general features of
integrability in the absence of boundaries, and consider as an example
the sine-Gordon model.  See Zamolodchikov and Zamolodchikov \cite{ZZ}
for a much more detailed review.

\subsection{Generalities}

A QFT is {\bf integrable} if it has an infinite set of mutually
commuting local integrals of motion (IMs).  According to the
Coleman-Mandula theorem \cite{CM}, an integrable, Lorentz-invariant
QFT in $D$ spatial dimensions has a trivial $S$ matrix, unless $D=1$.
Therefore, I henceforth restrict to 1 spatial dimension, with
coordinate $x$.  In this Section, I assume that there is no spatial
boundary; i.e., the theory is defined on the line $-\infty < x <
\infty$.

A trivial example of an integrable QFT is the theory of a free massive 
scalar field $\phi(x,t)$, with Lagrangian density
\be
\mathcal{L} = {1\over 2} \partial_{\mu}\phi \partial^{\mu} \phi 
- {1\over 2} m^{2} \phi^{2} \,.
\ee
Consider the following integrals of local densities \cite{Ku,KF}
\be
I_{2n} &=& \int_{-\infty}^{\infty}dx\ :\left[ 
{1\over 2}(\partial_{t} \partial_{x}^{n}\phi)^{2} +
{1\over 2}(\partial_{x}^{n+1}\phi)^{2} +
{1\over 2}m^{2}(\partial_{x}^{n}\phi)^{2} \right]:
\,, \non \\
I_{2n+1} &=&  \int_{-\infty}^{\infty}dx\ : \partial_{t} \phi\  
\partial_{x}^{2n+1}\phi :
\,, \qquad 
 n = 0 \,, 1 \,, 2 \,, \ldots 
\ee 
Using the equation of motion $\partial_{t}^{2}\phi = 
(\partial_{x}^{2} - m^{2}) \phi$, one can check that these quantities 
are conserved
${d\over dt} I_{n} = 0$,  $n = 0 \,, 1 \,, 2 \,, \ldots$, 
and are mutually commuting $\left[ I_{n} \,, I_{m} \right] = 0$.
Since $I_{0}$ is the energy and  $I_{1}$ is the momentum, the 
quantities $I_{n}$ are evidently higher-derivative generalizations of 
energy and momentum.

Since this is a free field theory, it is not surprising that it has
infinitely many IMs.  What is remarkable is that there do exist
interacting field theories, such as the sine-Gordon model, that are
integrable.  An important consequence of integrability is that the
multiparticle $S$ matrix can be factorized into a product of
two-particle $S$ matrices, as if the scattering occurred by a series
of {\bf elastic} two-particle collisions, the movement of the
particles in between being free.  ``Physicist proofs'' of this
fundamental result are given in Refs.  \cite{ZZ} and \cite{SW}.  An
important consistency condition for this factorization is known as the
Yang-Baxter Eq.  By solving this equation, and imposing the
constraints of crossing symmetry and unitarity, one can go a long way toward
determining the {\bf exact} two-particle (and hence, the full
multiparticle) $S$ matrix.

\subsection{The sine-Gordon model}

As an example, let us consider the so-called sine-Gordon model, which
is among the first known and most-studied integrable QFTs.  It is
convenient to go to Euclidean space $(x,y)$, with $z=x+ i y$, $\bar
z=x+ i y$.  The Lagrangian density is
\be
\mathcal{L}_{0} = 2 \partial_{z} \varphi \partial_{\bar z} \varphi - 
{m^{2}\over \beta^{2}} \cos (\beta \varphi) \,,
\label{bulkSG}
\ee
where $\varphi(z, \bar z)$ is a real scalar field. Since $\varphi$ is 
dimensionless (the number of spacetime dimensions being two), the 
coupling constant $\beta$ is also dimensionless. 

This model is known to be integrable at both the classical and quantum
levels.  To streamline the classical analysis, it is convenient to
eliminate $\beta$ from the field equation by rescaling the field,
i.e., by defining $\phi = \beta \varphi$; and to fix the mass
parameter by setting $m=2$.  This model has infinitely many conserved
currents \cite{Ku,KF,FT}
\be
\partial_{\bar z} T_{s+1} = \partial_{z} \Theta_{s-1} \,, 
\qquad 
\partial_{z} \overline{T}_{s+1} 
= \partial_{\bar z} \overline{\Theta}_{s-1} \,, \qquad s = 1 \,, 3 
\,, \ldots \,,
\label{continuity}
\ee 
starting with
\be
T_{2} &=& (\partial_{z}\phi)^{2} \,, \qquad \qquad \qquad \quad 
\Theta_{0} = -2 \cos \phi \,, \non  \\
T_{4} &=& (\partial_{z}^{2}\phi)^{2} - {1\over 4}(\partial_{z}\phi)^{4}
 \,, \qquad
\Theta_{2} = (\partial_{z}\phi)^{2} \cos \phi \,, \label{currents}
\ee
and the corresponding barred quantities are obtained by complex
conjugation.  It follows from (\ref{continuity}) that the local charges
\be
P_{s}= \int_{-\infty}^{\infty}dx\ (T_{s+1} +  \Theta_{s-1})  \,,
\qquad
\overline{P}_{s} = \int_{-\infty}^{\infty}dx\ 
(\overline{T}_{s+1} + \overline{\Theta}_{s-1})  \,,
\label{iom}
\ee
are conserved
\be
{d\over dy} P_{s}= 0 = {d\over dy} \overline{P}_{s} \,, 
\qquad s = 1 \,, 3 
\,, \ldots \,.
\ee
The energy and momentum are given by $P_{1} + \overline{P}_{1}$ and 
$P_{1} - \overline{P}_{1}$, respectively. The charges with $s  \ge 3$ 
are nontrivial -- their existence proves the classical integrability 
of the model.

The classical spectrum includes solitons and antisolitons. Indeed, 
the classical potential $V(\varphi) = - {m^{2}\over \beta^{2}} \cos 
(\beta \varphi)$ evidently has minima at $\varphi = 0$ (mod ${2\pi\over 
\beta}$ ). There exist stable finite-energy solutions of the classical 
field equations which interpolate between neighboring minima. 
Defining the topological charge
\be
T = {\beta\over 2 \pi}  \int_{-\infty}^{\infty}dx\ \partial_{x}\varphi
= {\beta\over 2 \pi} \left[ \varphi(x=\infty\,, y) -   
\varphi(x=-\infty\,, y) \right] \,,
\ee
the solutions with $T=+1$ and $T=-1$ are called solitons and 
antisolitons, respectively. There are also solutions with $T=0$ 
called breathers.

The quantum sine-Gordon model has particle-like states corresponding 
to these classical solutions, for $0 < \beta^{2} < 8\pi$. The exact 
two-particle $S$ matrix is given in \cite{ZZ}.

\section{Integrability in the presence of a 
boundary}\label{sec:boundary}

Following \cite{GZ}, let us now consider what happens when a spatial
boundary is introduced.  That is, I consider an integrable QFT on the
half line $-\infty < x \le 0$, which evidently has a boundary at
$x=0$.  The first problem to be addressed is to determine the boundary
conditions which preserve integrability.  Another important problem is
to determine the so-called boundary $S$ matrix, which describes
scattering off the boundary.  Integrability implies that particles
reflect elastically from the boundary, and that the boundary $S$
matrix obeys a boundary generalization \cite{Ch} of the Yang-Baxter
Eq.

Turning again to the example of the sine-Gordon model, let us consider
the Lagrangian \cite{GZ}
\be
L = \int_{-\infty}^{0}dx\ \left( 
2 \partial_{z} \phi \partial_{\bar z} \phi - 
4 \cos \phi \right) + B(\phi) \Big\vert_{x=0} \,. \label{boundSG1}
\ee
The Lagrangian density is essentially (\ref{bulkSG}) with the coupling
constant scaled away and with $m=2$.  The boundary potential $B(\phi)$
does not change the bulk equation of motion, but does affect the
boundary condition, which follows from the variational principle,
\be
\left( \partial_{x}\phi + {\partial B\over \partial \phi} 
\right)\Big\vert_{x=0} = 0 \,. \label{BC}
\ee

The question is: which $B(\phi)$ (if any) preserves integrability?
Clearly, the corresponding boundary conditions must be compatible with
some nontrivial IMs.  Since the boundary breaks translational
invariance, momentum-like quantities $\sim P_{s} - \overline{P}_{s}$
cannot be conserved.  The only hope is for energy-like quantities
$\sim P_{s} + \overline{P}_{s}$ to be conserved. Hence, consider
the quantity
\be
H_{s}= \int_{-\infty}^{0}dx\ \left[
(T_{s+1} +  \Theta_{s-1})  +
(\overline{T}_{s+1} + \overline{\Theta}_{s-1})  \right]  \,.
\ee
Computing the ``time'' derivative,
\be
{d\over dy} H_{s} &=& \int_{-\infty}^{0}dx\ 
\partial_{y} \left[ \quad  \right]
= \int_{-\infty}^{0}dx\ 
i\partial_{x} \left( T_{s+1} -  \Theta_{s-1}  -
\overline{T}_{s+1} + \overline{\Theta}_{s-1} \right) \non \\
&=& i \left( T_{s+1} -  \Theta_{s-1}  -
\overline{T}_{s+1} + \overline{\Theta}_{s-1} \right) \Big\vert_{x=0} 
\equiv i {d\over dy} \Sigma_{s} \,, \label{demand}
\ee
where the second equality on the first line follows from current
conservation (\ref{continuity}).  Hence, if there exists a quantity
$\Sigma_{s}$ obeying (\ref{demand}), then $H_{s} - i \Sigma_{s}$ is an
IM. That is, the boundary potential $B(\phi)$ should be chosen so that
(considering the first nontrivial case, $s=3$)
\be
\left( T_{4} -  \Theta_{2}  -
\overline{T}_{4} + \overline{\Theta}_{2} \right) \Big\vert_{x=0} 
= {d\over dy} \Sigma_{3} \,. \label{constraint}
\ee

Ghoshal and Zamolodchikov \cite{GZ} solved the constraint
(\ref{constraint}) for the boundary potential, and obtained the
result
\be
B(\phi) = 2 \alpha \cos ({1\over 2}(\phi - \phi_{0})) \,,
\label{boundSG2}
\ee
where $\alpha$ and $\phi_{0}$ are arbitrary boundary parameters.  The
model (\ref{boundSG1}), (\ref{boundSG2}) is known as the boundary
sine-Gordon model.  The corresponding boundary condition (\ref{BC}),
which reads $\left( \partial_{x}\phi - \alpha \sin ({1\over 2}(\phi -
\phi_{0})) \right)\Big\vert_{x=0} = 0$, interpolates between Neumann
($\alpha = 0$) and Dirichlet ($\alpha \rightarrow \infty$) boundary
conditions.  Ghoshal and Zamolodchikov conjectured that this model is
integrable at both the classical and quantum levels, and proposed the
boundary $S$ matrix for the solitons \cite{GZ} and breathers
\cite{Gh}.  Because there are (two) boundary parameters, the boundary
$S$ matrix exhibits a rich boundary boundstate structure
\cite{MD,BPTT}.

\section{Integrability and $N=1$ supersymmetry in the presence of a 
boundary}\label{sec:N1susy}

The ``bulk'' sine-Gordon model (\ref{bulkSG}) has a supersymmetric
generalization, the so-called $N=1$ sine-Gordon model \cite{ssg1}
\be
\mathcal{L}_{0} = 2 \left( \partial_{z}\phi \partial_{\bar z} \phi  
-  \bar \psi \partial_{z} \bar \psi 
+  \psi \partial_{\bar z} \psi 
- 2 \cos \phi - 2 \bar \psi \psi \cos {\phi\over 2} \right)\,,
\label{bulkSSG}
\ee
where $\psi$ and $\bar\psi$ are components of a Majorana Fermion field.  
(Again, the dimensionless bulk coupling constant has been scaled away,
and the mass parameter has been fixed to $m=2$.)
Indeed, this model has conserved supercurrents
\be
\partial_{\bar z} T_{3\over 2} = \partial_{z} \Theta_{-{1\over 2}} \,, 
\qquad 
\partial_{z} \overline{T}_{3\over 2} 
= \partial_{\bar z} \overline{\Theta}_{-{1\over 2}} \,, 
\ee 
where
\be
T_{3\over 2} = \partial_{z}\phi \psi \,, \qquad
\Theta_{-{1\over 2}}= -2 \bar \psi \sin {\phi\over 2} \,;
\ee
and corresponding conserved supersymmetry charges
\be
P_{1\over 2} = \int_{-\infty}^{\infty}dx\ 
(T_{3\over 2} +  \Theta_{-{1\over 2}})  \,,
\qquad
\overline{P}_{1\over 2} = \int_{-\infty}^{\infty}dx\ 
(\overline{T}_{3\over 2} + \overline{\Theta}_{-{1\over 2}})  \,.
\ee
Moreover, this model has an infinite set of local integrals of motion
\cite{ssg2}
\be
{d\over dy} P_{s}= 0 = {d\over dy} \overline{P}_{s} \,, 
\qquad s = 1 \,, 3 \,, \ldots \,,
\ee
(the corresponding currents for $s=1\,, 3$ are generalizations 
of (\ref{currents}) with additional terms involving the Fermion field) 
and is therefore integrable. Bulk $S$ matrices have been 
proposed for the solitons \cite{ABL} and breathers \cite{SW,Ah}.

One finally arrives at the question: are there boundary interactions
which preserve both integrability and supersymmetry?  This question
was first addressed by Inami, Odake and Zhang, who proposed the
Lagrangian \cite{IOZ}
\be
L = \int_{-\infty}^{0}dx\ \mathcal{L}_{0} + 
B(\phi, \psi, \bar\psi) \Big\vert_{x=0} \,, \label{boundSSG0}
\ee
where $\mathcal{L}_{0}$ is given by (\ref{bulkSSG}), and $B(\phi,
\psi, \bar\psi)$ is a boundary potential.  Imposing both integrability
(as in Eq.  (\ref{constraint})) and supersymmetry, they found
that the boundary potential is fixed up to a sign,
\be
B(\phi, \psi, \bar\psi) = \pm \left( 8 \cos {\phi\over 2} + \bar\psi 
\psi \right) \,.
\ee
That is, unlike the nonsupersymmetric ($N=0$) boundary sine-Gordon
model (\ref{boundSG1}), (\ref{boundSG2}), this model has {\bf no}
boundary parameters.

This no-go result (namely, that the combined constraints of
integrability and supersymmetry do not allow any free parameters in
the boundary action) seemed to me aesthetically unsatisfactory
and even paradoxical \cite{paradox}.  Hence, I decided to revisit this
problem \cite{Ne1}.  My main idea was to introduce a Fermionic
boundary degree of freedom in the boundary action.  Indeed, the $N=1$
solitons have an Ising-type RSOS degree of freedom \cite{ABL}, and
Ghoshal and Zamolodchikov \cite{GZ} introduced such a Fermionic
boundary degree of freedom to describe the Ising model in a boundary
magnetic field.  Thus, instead of (\ref{boundSSG0}), I proposed the
Lagrangian
\be
L = \int_{-\infty}^{0}dx\ \mathcal{L}_{0} + 
\left[ \pm \bar \psi \psi + ia {d\over dy} a
- 2 f(\phi) a (\psi \mp \bar \psi) + B(\phi) \right]\Big\vert_{x=0} \,, 
\label{boundSSG1}
\ee
where $\mathcal{L}_{0}$ is given by (\ref{bulkSSG}), $a$ is a
Hermitian Fermionic boundary degree of freedom which anticommutes with
both $\psi$ and $\bar \psi$, and $B(\phi)$, $f(\phi)$ are boundary
potentials.  Imposing both integrability (as in Eq.
(\ref{constraint})) and supersymmetry ($\sim P_{1\over 2} \pm
\overline{P}_{1\over 2})$, one finds that the boundary potentials are
given by \cite{Ne1}
\be
B(\phi) = 2 \alpha \cos ({1\over 2}(\phi - \phi_{0})) \,, \qquad
f(\phi) = {\sqrt{C}\over 2}\sin ({1\over 4}(\phi - D)) \,,
\label{boundSSG2}
\ee
where $C$, $D$ are known functions of $\alpha$, $\phi_{0}$.  That is,
the $N=1$ boundary sine-Gordon model (\ref{boundSSG1}),
(\ref{boundSSG2}) has two arbitrary boundary parameters ($\alpha$,
$\phi_{0}$) -- the same as the $N=0$ model (\ref{boundSG1}),
(\ref{boundSG2})!  A boundary $S$ matrix for the $N=1$ boundary
sine-Gordon model, which also depends on two boundary parameters, was
subsequently proposed by Bajnok et {\it al.} \cite{BPT}.

\section{Integrability and $N=2$ supersymmetry in the presence of a 
boundary}\label{sec:N2susy}

Encouraged by these results, I then decided to consider the $N=2$ case
\cite{Ne2}.  Indeed, the ``bulk'' sine-Gordon model (\ref{bulkSG})
also has an $N=2$ supersymmetric generalization \cite{KU1}
\be
\mathcal{L}_{0} &=&
{1\over 2}\big(-\partial_{z}\varphi^{-} \partial_{\bar z} \varphi^{+}
-\partial_{\bar z}\varphi^{-} \partial_{z} \varphi^{+} 
+ \bar \psi^{-} \partial_{z} \bar \psi^{+} 
+ \psi^{-} \partial_{\bar z} \psi^{+} 
+ \bar \psi^{+} \partial_{z} \bar \psi^{-}  \non \\
&+& \psi^{+} \partial_{\bar z} \psi^{-} \big) 
+ g \cos \varphi^{+} \bar \psi^{-} \psi^{-} +
g \cos \varphi^{-} \bar \psi^{+} \psi^{+}
+ g^{2} \sin \varphi^{+} \sin \varphi^{-} \,, \quad
\label{bulkN2SSG1}
\ee
where $\varphi^{\pm}$ form a complex scalar field; $\psi^{\pm}$ and
$\bar \psi^{\pm}$ are the components of a complex Dirac Fermion field;
$g$ is the bulk mass parameter; and here $z={1\over 2}(y+ix)$, $\bar
z={1\over 2}(y-ix)$.  (Again, the dimensionless bulk coupling constant
has been scaled away.)

This model has conserved supercurrents
\be
\partial_{\bar z} T_{3\over 2}^{\pm} 
= \partial_{z} \Theta_{-{1\over 2}}^{\pm} \,, 
\qquad 
\partial_{z} \overline{T}_{3\over 2}^{\pm} 
= \partial_{\bar z} \overline{\Theta}_{-{1\over 2}}^{\pm} \,, 
\ee 
where
\be
T_{3\over 2}^{\pm} = \partial_{z}\varphi^{\pm} \psi^{\pm} \,, \qquad
\Theta_{-{1\over 2}}^{\pm}= g \bar \psi^{\mp} \sin \varphi^{\pm} \,;
\ee
and corresponding conserved supersymmetry charges
\be
P_{1\over 2}^{\pm} = \int_{-\infty}^{\infty}dx\ 
(T_{3\over 2}^{\pm} -  \Theta_{-{1\over 2}}^{\pm})  \,,
\qquad
\overline{P}_{1\over 2}^{\pm} = \int_{-\infty}^{\infty}dx\ 
(\overline{T}_{3\over 2}^{\pm} 
- \overline{\Theta}_{-{1\over 2}}^{\pm})  \,.
\ee
Moreover, this model has an infinite set of local integrals of motion
\be
{d\over dy} P_{s}= 0 = {d\over dy} \overline{P}_{s} \,, 
\qquad s = 1 \,, 3 \,, \ldots \,,
\ee
and is therefore integrable. The bulk $S$ matrix was proposed in 
\cite{KU2}.

This model can be formulated on the half-line most simply when 
the bulk mass vanishes ($g=0$), in which case a suitable 
Lagrangian is \cite{Ne2} (see also \cite{Wa})
\be
\lefteqn{
L = \int_{-\infty}^{0}dx\ \mathcal{L}_{0} + \Big[
-{i\over 2}(\bar \psi^{+} \psi^{-}+\bar \psi^{-} \psi^{+})
-{1\over 2} a^{-} {d \over dy} a^{+}
- B(\varphi^{+} \,, \varphi^{-})}   \non \\ 
&&+{1\over 2} f^{+}(\varphi^{+}) a^{+} 
(\psi^{-} + \bar \psi^{-})
+ {1\over 2} f^{-}(\varphi^{-}) a^{-} 
(\psi^{+} + \bar \psi^{+}) \Big]\Big\vert_{x=0} \,, \label{bulkN2SSG2}
\ee
where $\mathcal{L}_{0}$ is given by (\ref{bulkN2SSG1}), $a^{\pm}$ are
Fermionic boundary degrees of freedom which anticommute with
$\psi^{\pm}$ and $\bar \psi^{\pm}$, and $B(\varphi^{+} \,,
\varphi^{-})$, $f^{\pm}(\varphi^{\pm})$ are boundary potentials.
Imposing both integrability (as in Eq.  (\ref{constraint})) and
supersymmetry ($\sim P_{1\over 2}^{\pm} + \overline{P}_{1\over
2}^{\pm})$, one finds that the boundary potentials are given 
by \cite{Ne2}
\be
B(\varphi^{+} \,, \varphi^{-}) &=& \alpha 
\cos ({1\over 2}(\varphi^{+} - \varphi^{+}_{0}))
\cos ({1\over 2}(\varphi^{-} - \varphi^{-}_{0})) \,, \non \\ 
f^{\pm}(\varphi^{\pm}) &=& 
{\sqrt{\alpha}\over 2}
\sin ({1\over 2}(\varphi^{\pm} - \varphi^{\pm}_{0})) \,.
\ee
Hence, there are three (!)  boundary parameters $\alpha$,
$\varphi^{\pm}_{0}$.  For the bulk massive case ($g \ne 0$), the
boundary action has more terms, and I have performed an analysis only
up to first order in $g$.  The boundary $S$ matrix for this model has
been discussed by Baseilhac and Koizumi \cite{BK}.  (See also
\cite{Wa,Ne3}.)

\section{Outlook}\label{sec:outlook}

Already for the the nonsupersymmetric ($N=0$) boundary sine-Gordon
model, there are many interesting questions that remain unanswered,
such as its relation to $\phi_{13}$-perturbed boundary minimal
conformal field theories (CFTs).  In the bulk case, it is known
\cite{restrict} that the $S$ matrices of the perturbed minimal models are
restrictions of the sine-Gordon $S$ matrix.  One would like to know if
something similar happens in the boundary case.

For the minimal models, the possible conformal boundary conditions
(CBCs) have been classified by Cardy \cite{Ca}.  (A CBC is
characterized in part by its boundary entropy \cite{AL}, similar to
the way that a bulk CFT is characterized by its central charge.)  The
boundary $S$ matrix of a perturbed CFT describes the boundary ``flow''
from one CBC to another.

Such issues (and more!)  can eventually also be addressed for the
$N=1$ and $N=2$ boundary sine-Gordon models which have been discussed 
here.

This work has also recently led to progress in formulating 
superconformal boundary Liouville models.  As is well known, the
Liouville model is closely related to the sine-Gordon model.  It is
conformal invariant, not just integrable.  For the $N=1$ and $N=2$
boundary Liouville models, the same Ans\"atze (\ref{boundSSG1}),
(\ref{bulkN2SSG2}) give one-parameter families of boundary actions
which are $N=1$ \cite{FH,ARS} and $N=2$ \cite{AY} superconformal
invariant, respectively.

\bigskip

{\it Dedicated to Stanley Deser, with gratitude for his guidance and support, 
and with best wishes.}

\end{document}